\documentclass[%
 aip,
rsi,
amsmath,amssymb,
 reprint,%
]{revtex4-1}

\usepackage{graphicx}% Include figure files
\usepackage{dcolumn}% Align table columns on decimal point
\usepackage{bm}% bold math
\usepackage[utf8]{inputenc}
\usepackage[T1]{fontenc}
\usepackage{mathptmx}
\usepackage{etoolbox}
\usepackage{booktabs}
\usepackage{multirow}
\renewcommand{\arraystretch}{1.2}

\usepackage{braket}
\usepackage{textcomp}
\usepackage{xcolor}
\usepackage{ulem}

\makeatletter
\def\@email#1#2{%
 \endgroup
 \patchcmd{\titleblock@produce}
  {\frontmatter@RRAPformat}
  {\frontmatter@RRAPformat{\produce@RRAP{*#1\href{mailto:#2}{#2}}}\frontmatter@RRAPformat}
  {}{}
}%
\makeatother
\begin{document}

% \preprint{AIP/123-QED}

\title{An agile radio-frequency source using internal linear sweeps of a direct digital synthesizer}

\author{Ethan Huegler}
\affiliation{ 
Department of Computer Science, University of Maryland, College Park, MD 20742, USA
}%

\author{Joshua C. Hill}%
\affiliation{%
DEVCOM Army Research Laboratory, 2800 Powder Mill Rd, Adelphi, MD 20783, USA
}%

\author{David H. Meyer}%
\email{david.h.meyer3.civ@army.mil}
\affiliation{%
DEVCOM Army Research Laboratory, 2800 Powder Mill Rd, Adelphi, MD 20783, USA
}%

\date{\today}% It is always \today, today,

\begin{abstract}
Agile rf sources are a common requirement for control systems in quantum science and technology platforms.
The direct digital synthesizer (DDS) often fills this role by allowing programmable control of the rf signals.
Due to limitations of the DDS architecture, implementing an agile rf source requires rapid and precisely-timed programming of discrete updates that restrict the source's agility. 
Here, we describe a microcontoller-based interface that exploits the DDS's internal linear sweep accumulator to perform both sequential linear sweeps, and standard discrete updates, at the $\sim10$\,\textmu s scale.
This allows updates to the swept parameter as fast as every 8\,ns with greatly reduced communication and memory overhead.
We demonstrate the utility of this system by using it as the reference to an optical phase-locked-loop to implement rapid, adjustable laser frequency sweeps in a Rydberg Electromagnetically Induced Transparency spectroscopy measurement.
\end{abstract}

\maketitle

\section{Introduction}

Quantum information science \& technology requires agile radio-frequency (rf) sources to satisfy the many demands of quantum control: for direct application to quantum systems, as drivers for acousto-optic modulators (AOMs), electro-optic modulators (EOMs), or inputs to various phase-locked-loops (PLLs).
These basic tools find applications in many different quantum platforms, including  trapped ions,\cite{paul_electromagnetic_1990} neutral-atom Bose-Einstein condensates,\cite{ketterle_evaporative_1996}, superconducting qubits,\cite{bardin_microwaves_2021} solid-state defect centers\cite{craik_microwave-assisted_2020} and atomic clocks.\cite{townes_atomic_2004}
As control systems scale up to satisfy more challenging applications, the required agile rf sources also scale in number, making cost and availability important metrics along with their performance.

Direct digital synthesizers (DDSs) are commonly used in these roles as they satisfy many of the desired characteristics.\cite{cordesses_direct_2004-1}
For example, the Analog Devices AD9959\cite{disclaimer} is a DDS that is widely used thanks to its four phase-synchronous outputs, where each has individually programmable amplitude, phase, and frequency.\cite{ad9959datasheet}
Moreover, it incorporates an internal linear sweep capability that can dynamically vary an output parameter.
These features allow for a highly flexible rf source that scales well with control system size.

Arbitrary and agile waveforms, including non-sinusoidal waveforms, required in experimental systems are often more difficult to implement with a DDS than other purpose-built technologies such as arbitrary waveform generators.
Beyond arbitrary outputs, the agility required (namely changes on a timescale faster than relevant dynamics being investigated) must remain synchronous with a larger experimental control system.
A DDS can implement agile waveforms via rapid changes to the rf amplitude, phase, and/or frequency of its sinusoidal output.
However, because the DDS's local memory only stores the current and next instruction, changing the output rapidly and synchronously with a larger control system requires frequent, precise, fast communications.

Current methods to solve this challenge typically use microcontollers or field-programmable-gate-arrays (FPGAs) to rapidly program the DDS outputs point-by-point from a stored memory of instructions based on external triggers.\cite{chen_2012,li_2016,du_2017}
This method allows for arbitrarily varying amplitude, frequency, and phase waveforms. However, it has limited time resolution because the finite communication speeds involved result in each update requiring $\gtrsim1$\,\textmu s.

Here, we demonstrate an alternative method of controlling an AD9959 DDS that circumvents this limitation, leading to a more agile rf source.
Similar to the existing methods, we employ a microcontroller to program the DDS.
However, instead of programming successive static outputs individually, we can also employ the linear sweep functionality of the DDS itself.
By successively programming new linear sweeps, we can generate an agile waveform that is synchronous with external triggers.
Thanks to reduced communication overhead to the DDS, this technique allows for much finer time resolution (down to $\sim8$\,ns) when sweeping a single parameter. The technique maintains the ability to statically adjust all parameters at the DDS programming timescale ($\sim 10$\,\textmu s).
Though this is insufficient for truly arbitrary waveform generation,
it significantly broadens the range of applications to which the DDS can be applied.
Furthermore, additional control can be obtained via augmentation with external voltage-controlled attenuators or phase shifters.

This work describes the hardware and firmware necessary for implementing what we have dubbed the DDS Sweeper.
We first provide a brief overview of DDS operational principles and limitations.
We then introduce the hardware used to implement the DDS Sweeper followed by an overview of the custom microcontroller firmware that provides an interface between a computer and the DDS.
The open-source code and pre-compiled binary of the firmware are available online.\cite{dds_sweeper}
We demonstrate the operation of the DDS Sweeper via simultaneous measurement of the amplitude, phase, and frequency of the outputs for various test waveform patterns.
Finally, we provide an example of the DDS sweeper's utility by using it as the frequency reference to an optical PLL in a Rydberg Electromagnetically Induced Transparency (EIT) spectroscopy measurement. We use the sweeper to enable rapid successive optical frequency sweeps of varying rates within a single measurement. 

\section{DDS Operation}
    
A DDS produces an output through the use of a phase accumulator and functions as a programmable fractional-frequency divider of an externally provided reference clock frequency.\cite{vankka2001} A phase accumulator stores two values, the current phase and the phase increment value. On each clock cycle (derived from the external reference), the phase increment is added to the current phase value. The phase and increment values are stored in 32 bit registers, and the operation depends on the modulo $2^{32}$ overflowing when adding 32 bit integers. The current phase value is passed through a phase to amplitude converter (typically a sine lookup table) and that amplitude is used as the input for a digital to analog converter (DAC). The filtered output of the DAC will be a sine wave of the frequency which has been programmed into the DDS.
For the detailed description of DDS operation that follows, we use the AD9959 as a concrete example. Further details beyond this summary can be found in the AD9959 datasheet.\cite{ad9959datasheet}
    
To program a frequency into the DDS, that phase increment -- or frequency tuning word ($FTW$) for the AD9959 -- must be provided. The $FTW$ can be calculated with $FTW = \frac{f_{out} \cdot 2^{32}}{f_{sys}}$ where $f_{out}$ is the desired output frequency and $f_{sys}$ is the system clock frequency of the DDS.
By nature of its operation, the DDS system clock frequency should be as high as possible to prevent discretization error.
It should also have low phase noise to limit the rf output phase noise.
For the AD9959, the maximum system clock frequency is 500\,MHz, which can either be provided directly, or via a programmable PLL multiplier from a lower frequency at the expense of higher output phase noise.

\begin{figure}[bt]
    \includegraphics[width=0.8\linewidth]{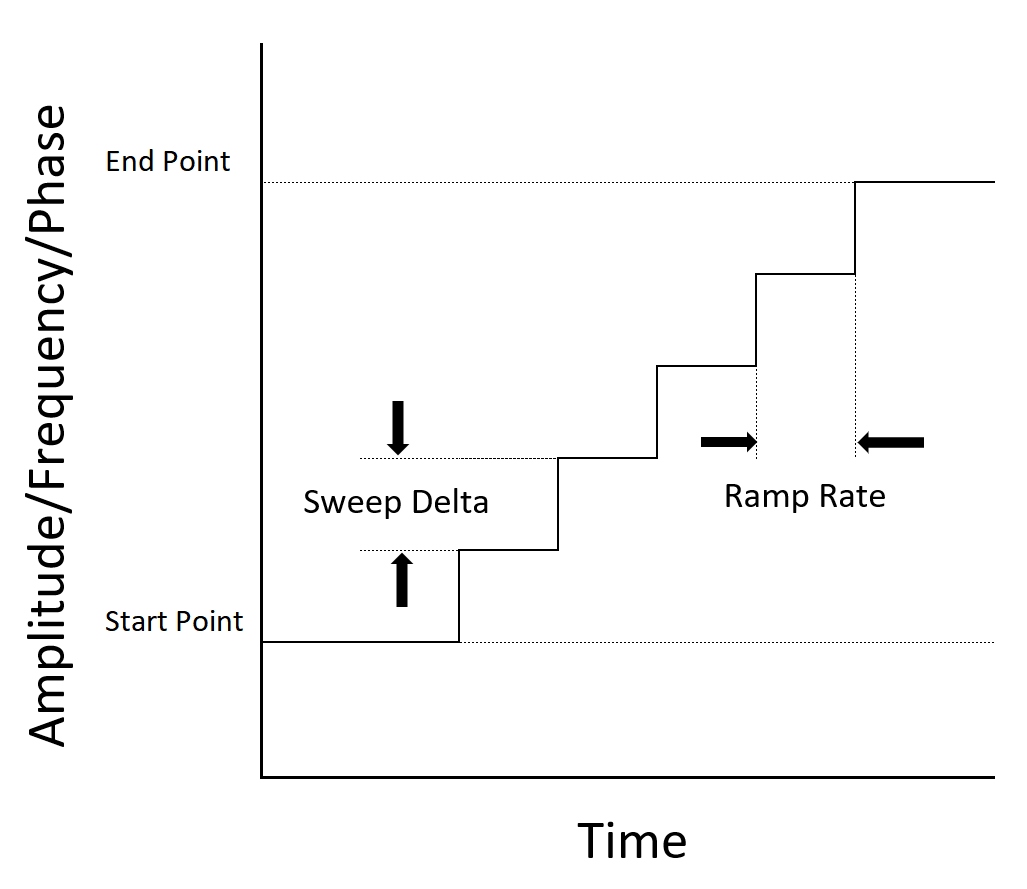}
    \caption{\label{fig:sweep-setup} Definitions of the parameters that define a generalized linear sweep for a DDS: Sweep Delta, Ramp Rate, Start Point, and End Point. }
\end{figure}

The DDS also has the ability to control the phase and amplitude. There is a 14-bit phase offset register which stores a phase offset word ($POW$) which is added to the phase accumulator value before the phase to amplitude converter,
allowing offsets spanning 0 to $360^\circ$.
$POW$ can be found by $POW = \frac{\phi \cdot 2^{14}}{360}$ where $\phi$ is the desired phase offset in degrees.

The amplitude is controlled by a 10-bit amplitude scale factor ($ASF$) which sets the ratio of maximum possible output current to the desired output current. It can be found with $ASF = r_I \cdot 2^{10}$ where $r_I$ is the desired ratio of output to maximum current.

The AD9959 DDS has additional functionality for performing a linear sweep of the output frequency, phase, or amplitude through the use of a sweep accumulator. The sweep accumulator is made up of a current value register and an 8-bit ramp rate counter register.
When linear sweep mode is active, the current value register is added to the $FTW$, $POW$, or $ASF$, depending on the parameter being swept. Fig.~\ref{fig:sweep-setup} shows the four parameters that define such a sweep:
start point, end point, sweep delta $DW$, and ramp rate $RR$.
The start and end points set the limits of the sweep while the sweep delta and ramp rate control the magnitude and duration, respectively, for each update of the accumulator.
The start point, end point, and sweep delta register values are all calculated identically to the tuning word of the parameter being swept (i.e. $FTW$, $POW$, or $ASF$).
The ramp rate is an 8-bit integer that is defined as $RR = \Delta t f_{sync}$, where $f_{sync}$ is the DDS sync clock which runs at one quarter of the system clock.
The AD9959 also allows for distinct values of sweep delta and ramp rate depending on the direction of the sweep where going from start to end points is defined as rising, the opposite direction as falling.
A dedicated digital input (Profile Pin) controls the sweep direction.

While in linear sweep mode, the DDS decrements the ramp rate counter register on every cycle of the sync clock.
When the ramp rate counter reaches zero, the profile pin is checked.
If the profile pin is logic high (low), the rising (falling) delta word $DW$ is added to the current value register and the ramp rate counter register is reset to the rising (falling) ramp rate $RR$.
Once the output of the phase accumulator and the sweep accumulator add up to either the start or end point the output is held constant.

All of the control values -- $FTW$, $POW$, $ASF$, and other configuration options -- are stored in registers on the DDS which can be written to via a serial interface. The DDS is equipped with an input buffer which stores all the writes it receives over the serial interface until an IO update signal is received. Upon receiving an IO update signal (IOUD) all of the registers are updated at once with their new values.
Using a microcontroller to quickly write to this buffer and trigger updates allows the DDS to produce an agile waveform.

\begin{figure}[tb]
    \includegraphics[width=0.8\linewidth]{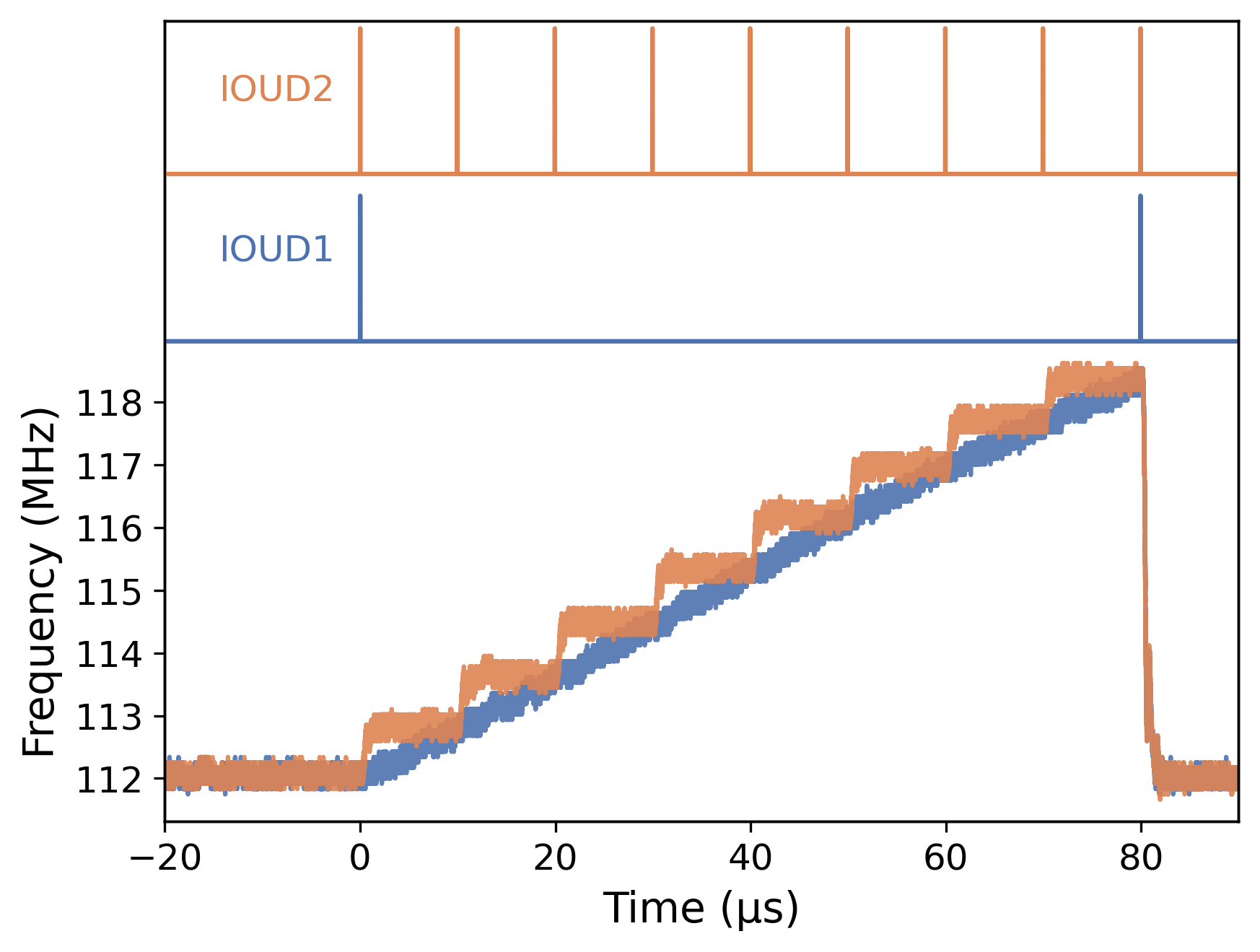}
    \caption{\label{fig:sweepvsstep} Comparative outputs of a frequency sweep using the DDS linear sweep mode (blue) versus the closest approximation using steps of the fastest stepping mode (orange). The IO update pulses for each sweep are also shown.}
\end{figure}

As mentioned in the Introduction, there are existing solutions which implement agile waveforms via rapid static updates (i.e. only changing $FTW$, $POW$, and $ASF$).\cite{chen_2012,li_2016,du_2017}
These solutions do not utilize the DDS's linear sweep capability,
largely because this mode can linearly sweep only a single parameter and is more challenging to implement than static updates.
However, it is still desirable to use the native linear sweeps of the DDS in many applications.
There are comparatively few examples in the literature that exploit these native linear sweeps.\cite{malek_2019} Those that do are often portions of a larger control system and lack detailed descriptions of their use or utility as we provide here.

Figure \ref{fig:sweepvsstep} highlights the differences between a static update sweep (orange) and a native linear sweep (blue) when generating a linearly ramping frequency that then resets to its initial value.
Using the native linear sweep, the DDS is only programmed twice (denoted by the corresponding IOUD trigger pulses), once for the sweep and once for the reset.
The static update sweep requires programming of the DDS at each step. Since the programming time is finite, the sweep itself is discretized.
By using the native linear sweep capability of the DDS, we can obtain higher quality linear sweeps with significantly lowered communication overhead between the microcontroller and the DDS itself.
Furthermore, since communication rates between a host computer and the microcontroller are often limited as well, the large table of pre-programmed values necessary for the static update sweep leads to long programming times ($\sim 1$\,s). This hinders rapid iteration of sweep parameters.
The reduced overhead of native linear sweeps can therefore improve overall experiment control sequence programming time.

\section{Hardware}

\begin{figure}[tb]
    \includegraphics[width=\linewidth]{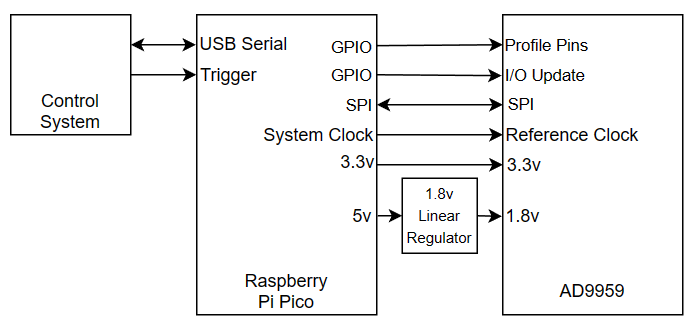}
    \caption{\label{fig:block-diagram} Block wiring diagram for the DDS Sweeper. A computer communicates with the Raspberry Pi Pico running the Sweeper's firmware via a USB serial interface. The Pico then communicates directly with the AD9959 via an SPI. It also provides power and an optional reference clock for the DDS.}
\end{figure}

The Sweeper is implemented using an Analog Devices 9959 (AD9959) evaluation board, and a Raspberry Pi Pico microcontroller board (based on the RP2040 microcontroller) to interface with a control system,\cite{picodatasheet} as seen in Fig.~\ref{fig:block-diagram}. 
The microcontroller is soldered to a custom interfacing PCB which provides the necessary routing and slots in the header pins of the DDS evaluation board. 
The DDS evaluation board also requires 3.3 V and 1.8 V power inputs, which the microcontroller can supply with the help of a linear regulator.
The DDS outputs are programmed by sending serial commands to the microcontroller via a USB interface.
The microcontroller interprets these commands and programs the appropriate registers of the DDS via a Serial Peripheral Interface (SPI) at 62.5 Mbits/s.\cite{leens_IEEE_2009}

The Raspberry Pi Pico was chosen as the microcontroller for its Direct Memory Access (DMA) channels and Programmable Input Output (PIO) cores. DMA channels are dedicated hardware for moving memory without utilizing the main core. We use the DMA to send the instruction table stored in memory to the SPI controller as quickly as possible while allowing the main core to simultaneously prepare the next instruction. The PIO cores are independent processor cores with limited functionality but direct access to the GPIO pins. The Pico's PIO cores allow the Sweeper to precisely time multiple aspects of its operation, including: programming of the registers via an SPI interface, the IO Update (IOUD) triggers, and the profile pins that control sweep directions.

In the default configuration, the microcontroller also provides its 125\,MHz system clock to the DDS as a reference clock. With the default PLL multiplier of 4 this gives the DDS a system clock of 500\,MHz, which is the maximum system clock supported by the AD9959. The Sweeper can also be configured to allow an external reference clock for the DDS.

\section{Firmware}

The Sweeper operates in two modes: manual mode or buffered execution mode. Buffered mode has three sub-modes: single steps, linear sweeps, or a combination of both

In manual mode, instructions sent to the Sweeper over the USB interface update the outputs of the DDS in real time.

In buffered execution mode, the Sweeper accepts a sequence of instructions which it stores in memory. At a later point the Sweeper can recall the sequence and successively program the instructions into the DDS.
To keep the outputs synchronized with a larger system, 
the buffered execution process can be triggered from an external source for each step of the sequence, or the microcontroller can time itself based on the Pico's clock.

The microcontroller writes each instruction of a sequence to the input buffer of the DDS immediately after the previous IO update has completed.
When the time for updated output arrives, the IO Update signal is sent to the DDS and the next instruction is written to the input buffer of the DDS.
This minimizes the delay between triggers and output updates.
When receiving external trigger signals, the Sweeper has a pipeline delay of 4 clock cycles, at 125 MHz that will be 32 ns $\pm 8$ ns, since the microcontroller buffers GPIO inputs to occur on clock cycle edges.

\begin{table}[b]
    \centering
    \caption{\label{tab:timing-clocks} Minimum time between instructions for each buffered execution mode.}
    \begin{tabular}{l c c c c c}
        &        & \multicolumn{4}{c}{Num. of Outputs} \\ \cmidrule(lr){3-6}
        &        & 1  & 2  & 3  & 4  \\ \toprule
\multirow{2}{*}{Single Stepping} & cycles & 500       & 750       & 1000      & 1250      \\
  & $\mu s$  & 4 & 6  & 8  & 10 \\ \cmidrule(lr){2-6}
\multirow{2}{*}{Sweep Mode} & cycles    & 1000      & 1500      & 2000      & 2500      \\
  & $\mu s$  & 8  & 12  & 16  & 20  \\\cmidrule(lr){2-6}
\multirow{2}{*}{Sweep and Step Mode} & cycles    & 1250      & 2000      & 2750      & 3500      \\
  & $\mu s$  & 12  & 18  & 24  & 30  \\\bottomrule
    \end{tabular}
\end{table}

A buffered sequence can be programmed in one of three ways: as single steps, a sequence of linear sweeps, or a combination of both.
Single stepping allows discrete changes to frequency, amplitude, and phase parameters simultaneously, replicating the mode of operation that others have implemented.
The sweep and combination modes utilize the DDS linear sweep functionality, but require setting more registers of the DDS and therefore require a longer dwell period in between instructions to send all the required bytes, as seen in Tab.~\ref{tab:timing-clocks}.
Similarly, the instructions for sweep and combined modes are longer so fewer of them can be stored in memory.
Instructions are only stored for the channels being utilized, and the maximum number of instructions for each mode of operation can be seen in Tab.~\ref{tab:instruction-count}.

Once the sequence is programmed, sending a start command to the Sweeper will begin sending the instructions stored in memory.
Since system memory will not persist through a power cycle, there is functionality to store and recover instruction sequences from non-volatile storage on the microcontroller board.

\begin{table}[tb]
    \centering
    \caption{\label{tab:instruction-count} Maximum number of stored instructions for each buffered execution mode.}
    \begin{tabular}{l c c c c}
       & \multicolumn{4}{c}{Num. of Outputs} \\ \cmidrule(lr){2-5}
        & 1  & 2  & 3  & 4  \\ \toprule
\multicolumn{5}{c}{Internally Timed}                                \\ \midrule
Single Stepping     & 5000      & 5000      & 5000      & 4032      \\
Sweep Mode          & 5000      & 3895      & 2611      & 1964      \\
Sweep and Step      & 5000      & 3148      & 2108      & 1584      \\ \midrule
\multicolumn{5}{c}{Externally Timed }                               \\ \midrule
Single Stepping     & 16656     & 8615      & 5810      & 4383      \\
Sweep Mode          & 8327      & 4234      & 2838      & 2135      \\
Sweep and Step      & 6751      & 3422      & 2291      & 1722      \\ \bottomrule
    \end{tabular}
\end{table}

Users send instructions over USB to the microcontroller with the desired outputs in units of Hertz, Degrees, and percentages. The microcontroller calculates the tuning words from those values and translates them into the expected bit alignment for the DDS. The tuning resolutions for frequency, phase, and amplitude are $0.022^\circ$, $0.116$\,Hz, and 0.1\% of maximum output current, respectively.

If it is desired for the Sweeper to time itself, an additional parameter can be sent with the number of clock cycles the instruction should take. At runtime, these wait lengths are sent to a PIO core through DMA to handle the timing, based on the Prawnblaster Psuedoclock project.\cite{prawnblaster}

\subsection{Linear Sweep Control}

Performing arbitrary sweeps on the DDS as part of a sequence requires great consideration of the internal operational details of the AD9959 DDS.

A reliable rising sweep depends upon the sweep accumulator being at zero when the sweep begins. If one sweep follows another, the sweep accumulator will not generally be at zero. The AD9959 does have an ``autoclear sweep accumulator'' functionality, which, when enabled, will reset the sweep accumulator to zero upon an IO Update signal. This does allow for running successive upward sweeps.

The AD9959 implementation of linear sweeps does not inherently support arbitrary falling sweeps.
This is the because the sweep accumulator is an unsigned register that is always interpreted as positive and therefore added to the start point value: there is no way to subtract the sweep accumulation from the start point.
Since the sweep delta is not applied if the current value is greater than the end point, the combined result is if the start point is larger than the end point the frequency output will remain constant at the start point value.

To run a falling sweep the start point must be programmed as the lowest output desired from the sweep, and the end point programmed as the highest output desired from the sweep. Then the sweep accumulator must first be filled up by a rising sweep so that when the profile pin is set to low the falling sweep delta can be subtracted from the sweep accumulator until the sweep accumulator holds a value of zero and the DDS output is equal to the value programmed into start point.

Since an arbitrary sequence cannot guarantee that all falling sweeps will be preceded by a rising sweep, we implement a hidden rising sweep before a falling sweep.
The effect of these hidden sweeps is minimized by setting the rising sweep delta to its maximum value,
then the sweep accumulator will be filled up by just one instance of the rising sweep delta being applied. If the rising ramp rate is set to 1, then it will only take one cycle of the sync clock (4 cycles of the system clock) to fill up the sweep accumulator (with the default 500\,MHz system clock this is 8\,ns). 
This works with the limitation that the falling ramp rate must be set to 1 to match the rising ramp rate that filled the accumulator.
This limitation was determined empirically as the AD9959 behavior in this circumstance is not well documented.

\begin{figure}[tb]
    \centering
    \includegraphics[width=0.8\linewidth]{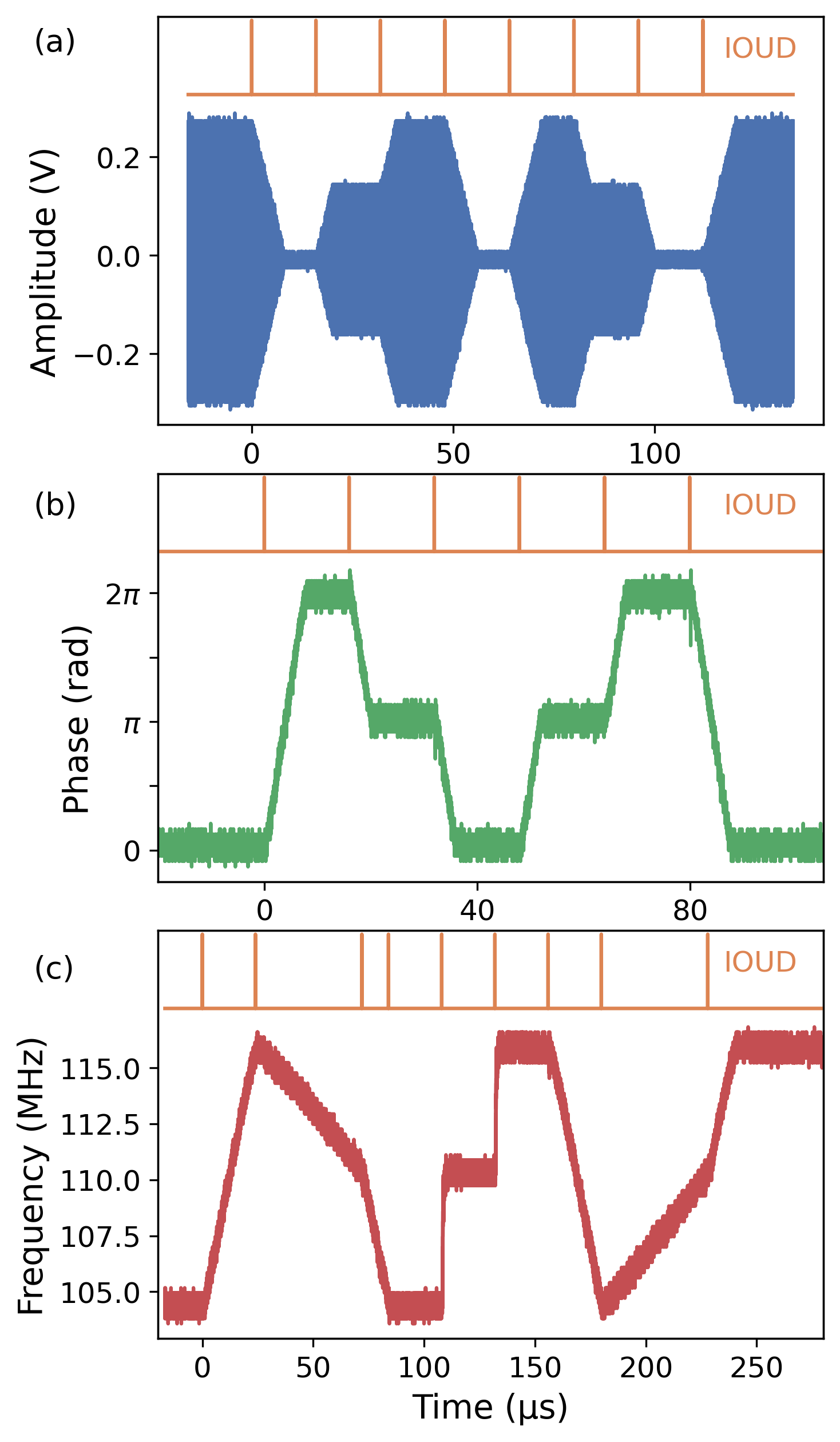}
    \caption{Demonstrative measurements of the DDS Sweeper capabilities showing (a) amplitude sweep, (b) phase sweep, (c) frequency sweep. Each trace also shows the digital IO update pulses marking the start of each new instruction to the AD9959.}
    \label{fig:sweeps}
\end{figure}

With this limitation in mind, the minimum sweep rates differ between rising and falling sweeps.
The maximum frequency sweep rate is $\pm62.5$\,GHz/s while the minimum sweep rates are 57\,kHz/s and $-14.5$\,MHz/s.
For the amplitude scale, the maximum rate is $\pm100$\%/8\,ns with minimum rates of 100\%/2.1\,ms and $-100\%$/8.2\,\textmu s.
The phase maximum rate is $\pm360^\circ$/2.1\,\textmu s with minimum rates of $360^\circ$/33.43\,ms and $-360^\circ$/133.3\,\textmu s.

An alternative solution to running downward sweeps is to turn the autoclear sweep accumulator functionality off for downward sweeps. Somehow this allows successive downward sweeps, but it makes the range of downward sweeps dependent on preceding upward sweep. If an upward sweep only fills the sweep accumulator half way, the next downward sweep will not be able to drain more than the half of the sweep accumulator, even if it was programmed to sweep over the full range of the accumulator.

\section{Example Outputs}

To confirm functionality of the sweeper, we implement independent measures of the amplitude, frequency, and phase of the output(s).
Due to the short timescales of the dynamic features of the DDS Sweeper, these measurements must be able to resolve similar timescales.
To measure the amplitude, we use an oscilloscope with a 1\,GHz bandwidth.
The relative phase between two channels of the DDS (one serving as a fixed reference) is measured using a phase frequency detector provided by a HMC439 evaluation board.\cite{disclaimer} The outputs of the detector are recorded by the same oscilloscope.
The frequency is measured using a delayed self-homodyne measurement. The output of the recombining mixer is low-pass filtered and measured directly by the same oscilloscope.
Delay line length and operating frequency shown in the plots were chosen to ensure the homodyne output was centered within the linear regime.
By appropriate power splitting and amplification, all three measurements could be performed simultaneously on a single output.

\begin{figure}[tb]
    \centering
    \includegraphics[width=0.8\linewidth]{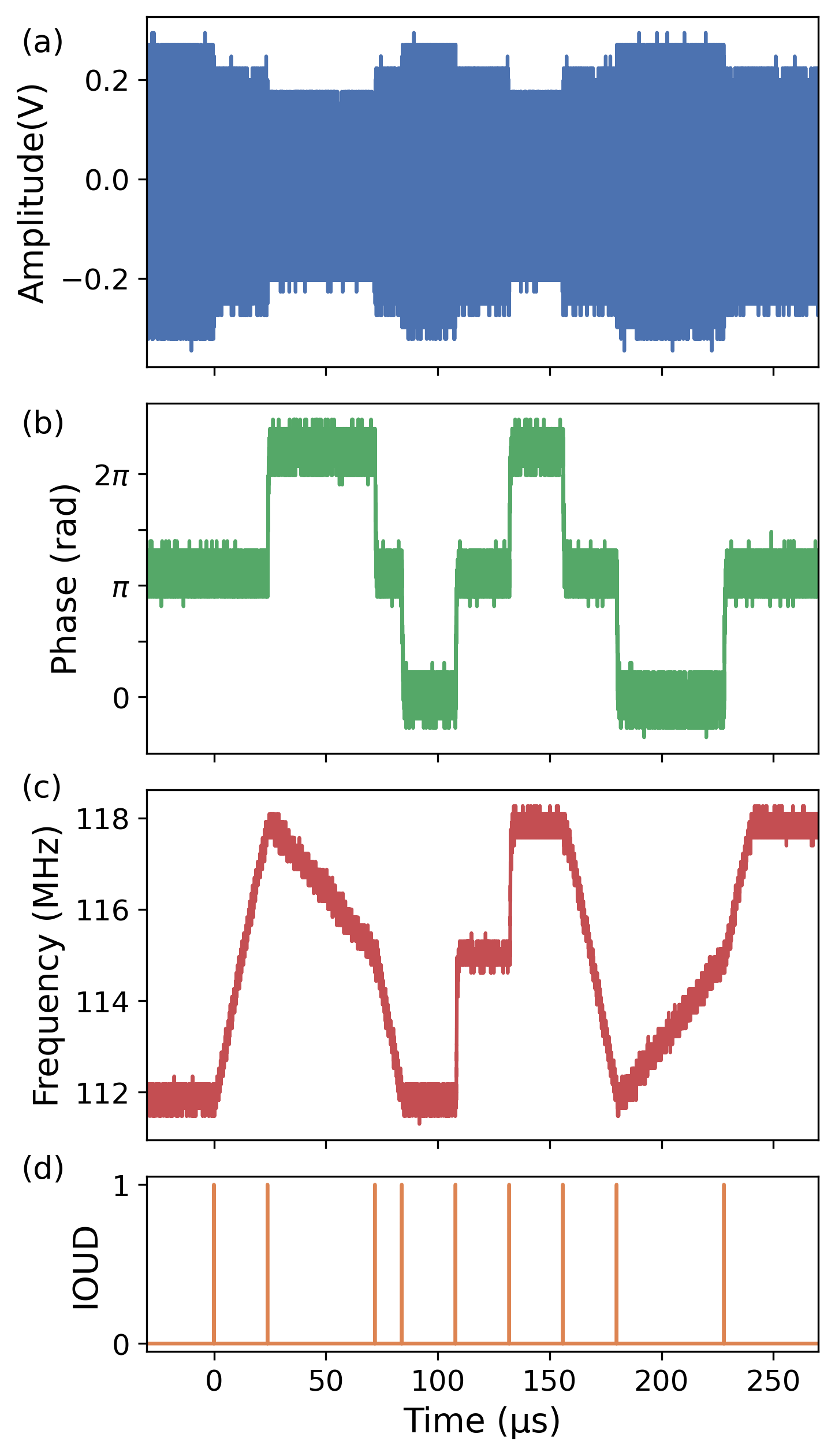}
    \caption{Demonstrative measurements of the DDS Sweeper capabilities showing (a) amplitude steps, (b) phase steps, and (c) frequency sweep performed simultaneously on a single output channel. Also shown is the (d) digital IO update pulses marking the start of each new instruction to the AD9959.}
    \label{fig:combinedsweeps}
\end{figure}

In Fig.~\ref{fig:sweeps} we show demonstrative successive sweeps of the individual parameters. Each sub-figure represents a single sweep on a distinct parameter. The sub-figures also show the update pulses that mark when the microcontroller sent a new instruction to the DDS.
For all of these sequences, the DDS Sweeper uses its own internal timing to determine the start of each instruction.

Fig.~\ref{fig:sweeps}(a) shows a measurement of a constant 100 MHz output from the DDS as the amplitude scale factor is swept.
In Fig.~\ref{fig:sweeps}(b), two outputs of the DDS are kept at a constant 100\,MHz. The output of channel 0 is held at a constant phase offset while the channel 1 output is run through a sequence of phase changes spanning 0 to $2\pi$.
Fig.~\ref{fig:sweeps}(c) shows a measurement of the output frequency.
Note that this sequence includes successive upward and downward linear sweeps with different sweep rates.
These sequences were chosen to have a mixture of linear sweeps and discrete jumps of the varying parameter in order to demonstrate the flexibility of the DDS Sweeper.

Fig.~\ref{fig:combinedsweeps} shows the Sweeper operating in Sweep and Step mode, where a single parameter can employ linear sweeps and the other parameter can be discretely changed at each update. Here the frequency is the parameter being swept (c) while amplitude (a) and phase offset (b) are stepped simultaneously. Part (d) shows the the update pulses from the microcontroller. This dynamic waveform only sends 9 instructions to the DDS, greatly reducing communication overhead.

\begin{figure}[tb]
    \centering
    \includegraphics[width=0.9\linewidth]{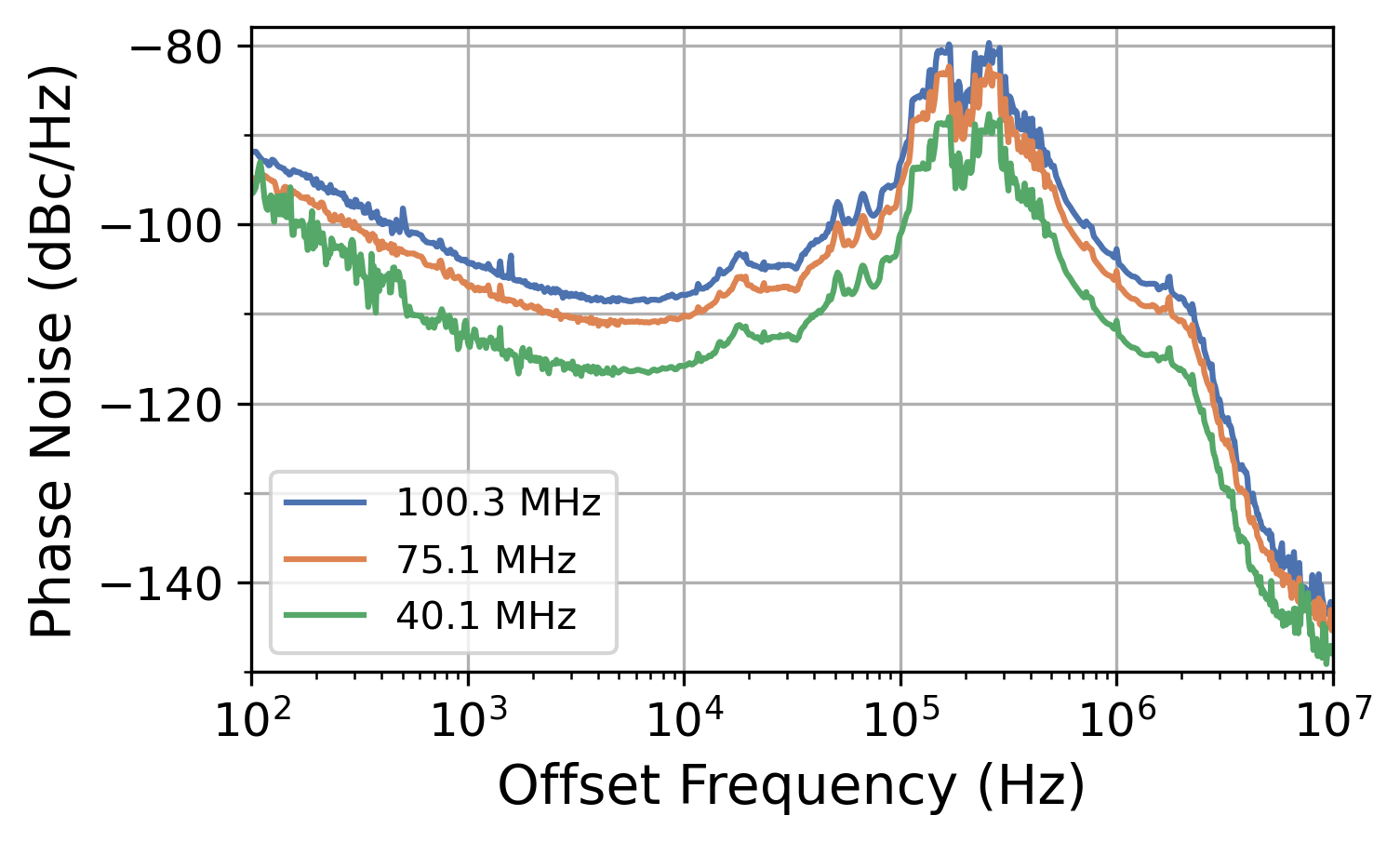}
    \caption{Phase noise of the DDS output when clocked by the Pico at 100.3, 75.1 and 40.1\,MHz output frequencies. The phase noise is dominated by the phase noise of the Pico's crystal oscillator from which the clock signal is derived.}
    \label{fig:phasenoise}
\end{figure}

In its default configuration, the DDS Sweeper derives the DDS reference clock from the Pico's on-board crystal oscillator. Because this reference is used directly to produce the outputs, it's phase noise will have a strong impact on the phase noise of the outputs. 
Using a Berkeley Nucleonics 7340 phase noise tester,\cite{disclaimer} we measured the phase noise of the DDS when clocked directly from the Pico at three frequencies: 100.3, 75.1, and 40.1\,MHz, as shown in Fig.~\ref{fig:phasenoise}.
The overall noise floor is approximately 20\,dB higher than the minimum noise of the DDS (when using the on-board PLL and multiplier) and the large peak around 200\,kHz from the PLL is more pronounced.
This is to be expected as the Pico's crystal oscillator is not designed to be highly performant.
If improved phase noise is required for a given application, the DDS Sweeper can be configured to allow for an externally-provided reference of the DDS.

\begin{figure*}[t]
    \centering
    \includegraphics[width=0.8\linewidth]{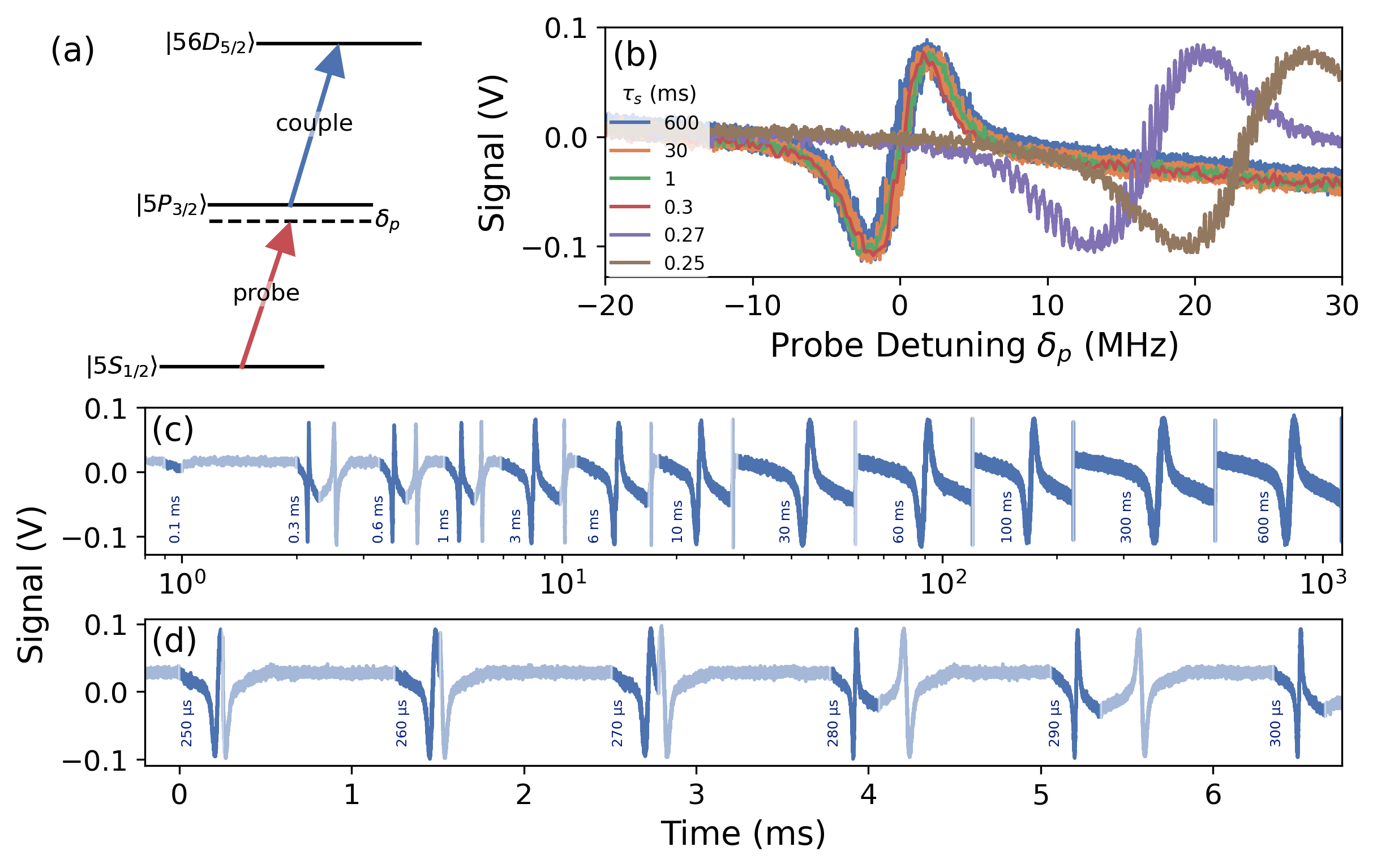}
    \caption{Example Rydberg EIT spectroscopy sweeps.
    (a) Level diagram for Rydberg EIT.
    The coupling laser is resonant with the excited transition, whereas the probe laser is swept over resonance ($\delta_p=0$) with the intermediate state transition.
    (b) Example Rydberg EIT probe sweeps for different sweep periods $\tau_s$.
    (c) Raw timetrace of the sweeps with periods spanning 0.1 to 600\,ms, taken in a single continuous measurement. The lightly shaded regions show the reset windows between each sweep. Each sweep is labelled by $\tau_s$.
    (d) Raw timetrace of sweeps with periods spanning 250 to 300\,\textmu s, taken in a single continuous measurement. Each sweep is labelled by $\tau_s$.}
    \label{fig:eitsweep}
\end{figure*}

\section{Example Usage}

Finally, we demonstrate the sweeper's utility by performing a rapid series of spectroscopy scans of varying duration.
This is accomplished by using a channel from the Sweeper as the reference for an optical Phase-Locked-Loop. This method of stabilization compares the rf beatnote between two lasers. A first (reference) laser is independently stabilized to atomic spectroscopy. The second (controlled) laser receives feedback from the optical PLL to stabilize the beatnote to the frequency set by the rf reference (DDS sweeper).
As the rf reference is swept, the controlled laser will be swept.
Using an agile rf reference allows for agile, precise changes in the control laser's frequency.

Our demonstration measures Rydberg Electromagnetically-Induced-Transparency (EIT) spectroscopy, using the experimental apparatus and technique described in Ref.~ \onlinecite{meyer_assessment_2020}.
This measurement involves a 780\,nm probe laser and a 480\,nm coupling laser that counter-propagate through a room-temperature vapor cell filled with natural isotopic abundance rubidium.
The probe laser nominally couples the $\ket{5S_{1/2}}$ ground state with the $^{85}$Rb $\ket{5P_{3/2}}$ first excited state.
Its frequency difference (detuning) relative to the non-Doppler-shifted atomic resonance at $\delta_p$=0 is controlled via the optical PLL described previously.
The coupling laser couples $\ket{5P_{3/2}}$ to the $\ket{56D_{5/2}}$ Rydberg state and is kept resonant with this transition.
When both lasers are resonant, EIT is established between the ground and Rydberg state, leading to reduced atomic absorption of the probe light.
See Figure \ref{fig:eitsweep}(a) for the level diagram.
By using an optical homodyne measurement of the transmitted probe light, we measure the corresponding phase shift of the probe field.
The photodiode output is recorded using a 50\,$\Omega$ terminated oscilloscope.

The goal of this demonstration is to empirically determine, in a single continuous measurement, how quickly the probe laser could be swept through resonance without introducing errors or distortion to the dispersive EIT signal.
To this end, we program the Sweeper to linearly sweep in frequency between two fixed points at variable rate, reset to the sweep start point, then dwell for 1\,ms to allow the probe laser frequency to settle before the next sweep.
For each sweep, the probe detuning spans from $-30$ to $30$\,MHz and the sweep rate is adjusted by changing the sweep duration. The optical PLL contains a x16 multiplier of the rf reference so the total sweep of the DDS output is 3.75\,MHz.

Figure \ref{fig:eitsweep}(c) shows a single output spectroscopic signal trace as the sweep period, $\tau_s$, was increased from 100\,\textmu s to 600\,ms at multiples of 1, 3, and 6.
The lightly shaded regions denote the 1\,ms reset window between each sweep where the reverse EIT signal can be seen as the optical PLL moves the probe frequency back to the sweep start at a finite rate due to the instantaneous frequency change.
From this measurement, it is clear that the critical sweep rate for which slower sweeps do not distort the signal corresponds to a period between 100 and 300\,\textmu s.

A second timetrace is shown in Figure \ref{fig:eitsweep}(d), with sweep times spanning 250 to 300\,\textmu s in steps of 10\,\textmu s.
These sweeps show the critical sweep rate corresponds to a period of approximately 270\,\textmu s or 0.22\,MHz/\textmu s.
From repeated measurements, we further determined that this critical rate is borderline distortionary to the signal on average, with random fluctuations in experimental parameters causing the measured critical period to fluctuate on the order of $\pm5$\,\textmu s.

Figure \ref{fig:eitsweep}(b) shows the sweeps with 600, 30, 1, 0.3, 0.27, and 0.25\,ms periods versus probe detuning $\delta_p$.
The apparent wider trace widths as the sweep time is increased is an artifact of each sweep being sampled at the same rate, which makes the longer sweeps more susceptible to electronic noise.
However, there is also evident a slight broadening of the dispersive feature as the sweep time is increased, attributable to systematic errors in the measurement at longer timescales (such as finite laser frequency stability or varying background electric fields).
As the sweep time is decreased, the optical PLL bandwidth is reached. This results in the EIT feature dragging away from resonance and observable local ringing in the frequency, as seen in the 270 and 250\,\textmu s sweep traces of Figure \ref{fig:eitsweep}(b).

\section{Conclusion}

We have described a custom firmware for the Raspberry Pi Pico microcontroller that can control an Analog Devices AD9959 four-channel DDS.
By exploiting the inherent linear sweep accumulators of the DDS, our implementation (the Sweeper) achieves agile rf outputs and fast updates that are synchronous with an external control system.
We also demonstrated various types of waveforms that can be produced. Moreover, by referencing an optical phase locked loop to the Sweeper, we highlighted a example application by performing spectroscopic sweeps of a Rydberg EIT signal.

The Sweeper is capable of filling an important role in any application where simple, agile, rf sources are ubiquitous. Quantum science experiments are a noteworthy example where demands for increased scale could make it advantageous compared to contemporary alternatives,
such as arbitrary waveform generators.
Beyond its performance, the device consists of readily-available cost-effective components
that have favorable size, weight, and power characteristics.
These features make the Sweeper a candidate for replacing or supplementing more expensive rf sources used in many experiments.

There are multiple potential means of enhancing the DDS Sweeper firmware.
The Sweeper currently uses the standard SPI protocol for communications between the Pico and the AD9959. This protocol could be modified to use a custom multi-channel SPI version (supported by the DDS) allowing up to a factor of four decrease in the time required to program the DDS at each update.
The serial interface between the control computer and the Pico could also be improved by using a more efficient character encoding than the standard ASCII we have implemented. This could decrease the time required to send instructions to the DDS Sweeper by approximately a factor of two.
Finally, the flash memory of the Pico board could be leveraged to increase the total number of instructions that can be programmed in a single sequence.

\begin{acknowledgments}
We wish to acknowledge Kevin Cox for helpful discussions.
EH recognizes financial support from the National Security Scholars Summer Internship Program (NSSSIP).

The views, opinions and/or findings expressed are those of the authors and should not be interpreted as representing the official views or policies of the Department of Defense or the U.S. Government.
\end{acknowledgments}

\section*{Author Declarations}

\subsection*{Conflict of Interest}

The authors have no conflicts to disclose.

\subsection*{Author Contributions}

\textbf{E. Huegler}: conceptualization (supporting); software (lead); investigation (lead); visualization (equal); writing -- original draft (equal); writing -- review and editing (equal)
\textbf{J. C. Hill}: investigation (supporting); writing -- review and editing (equal)
\textbf{D. H. Meyer}: conceptualization (lead); investigation (supporting); visualization (equal), writing -- original draft (equal); writing -- review and editing (equal)

\section*{Data Availability Statement}

The data presented are available from the corresponding author upon reasonable request.

%\nocite{*}
\bibliography{refs}% Produces the bibliography via BibTeX.

\begin{thebibliography}{18}
\bibitem [{dis()}]{disclaimer}%
  \BibitemOpen
  \href@noop {} {}\bibinfo {note} {References to commercial devices do not
  constitute an endorsement by the US Government or the Army Research
  Laboratory.}\BibitemShut {Stop}%
\end{thebibliography}

\end{document}